\documentstyle[prl,multicol,aps,epsf]{revtex}
\begin{document}
\draft
\title{Massless Dirac fermions, gauge fields, and underdoped cuprates}
\author{Don H. Kim, Patrick A. Lee, and Xiao-Gang Wen}
\address{Department of Physics, Massachussetts Institute of Technology,
Cambridge, MA, 02139}
\date{\today}
\maketitle
\begin{abstract}
We study 2+1 dimensional  massless
Dirac fermions and  bosons coupled to a U(1)
gauge field as a model for  underdoped cuprates.    
We find that the uniform susceptibility  and the specific 
heat coefficient are
logarithmically enhanced (compared to linear-in-$T$ behavior) due
to the fluctuation of transverse gauge field which is the only 
massless mode  at
finite boson density.  We analyze existing data, and find good 
agreement in the spin gap phase.  Within our picture, the drop of the 
susceptibility below the superconducting $T_c$ arises from the 
suppression of gauge fluctuations.
\end{abstract}

\begin{multicols}{2}
\tighten
 Recent experiments have  indicated
 the existence in the normal state underdoped cuprate superconductor
 of a gap with the same anisotropy as the $d$-wave
superconducting gap.
  One proposed explanation involves
 spin-charge separation: an electron in these
highly correlated materials is a composite 
object made of a spin $\frac{1}{2}$
neutral fermion (spinon) and a spinless charged boson (holon).  The 
suppression of normal state
 magnetic excitation seen in NMR and neutron scattering
is thus viewed as a singlet pairing of neutral fermions in the
absence of coherence among holons.  As a possible 
realization of this idea,
two of us have taken the $t$-$J$ model (which is believed to capture
essential physics of ${\rm CuO}_2$ planes) and developed a slave boson 
mean field theory\cite{wenlee} that 
extends the local SU(2) symmetry at 
half-filling to the finite concentration of holes by introducing a
SU(2) doublet of slave boson field.
Among the  mean field phases reported in Ref.\cite{wenlee} the
so-called staggered flux (sF) phase  
(which is connected to $d$-wave pairing phase
by a local SU(2) transformation) was argued to describe the
pseudogap in underdoped cuprates.  The low energy physics of
this phase can be described by massless Dirac fermions,
non-relativistic bosons, and a massless U(1) gauge field 
which together with
two massive gauge fields forms  SU(2) gauge fields that represent
the fluctuations around the mean field.

   The purpose of this paper is to address the low energy
effective theory of the sF phase as a U(1) gauge theory problem. 
 Although Dirac fermions coupled to a gauge field
had been considered in several contexts in the past\cite{marston}, 
we shall see that
interesting new physics emerges 
when massless Dirac fermions are coupled
to a gauge field that is also coupled 
to a compressible boson current.
More specifically, the Lorentz symmetry breaking due to coupling to
the bosons results in the renormalization of fermion velocity 
which have consequences on physical properties such as uniform 
susceptibility $\chi_u$ and
electronic specific heat $c_v^{el}$.  Experimentally, $\chi_u$ of 
underdoped cuprates begins to
decrease with lowering of temperature far above the superconducting
$T_c$\cite{nakano,taki,bankay}. 
 Electronic specific heat experiments\cite{loram,loramYBCO} show that 
$\gamma(T)$ $(\equiv c_v^{el}(T)/T)$  behaves quite similar
to $\chi_u$.  
Although constant Wilson ratio ($\gamma/\chi_u$) 
is a hallmark of Fermi liquid theory, 
the anomalous temperature dependence calls for a 
departure from the time-honored theory of most metals.
We make a case  that the puzzling normal state
behavior of $\chi_u$ and $\gamma$ may be viewed as
{\it enhancement} over linear-in-$T$ $\chi_u$ and $\gamma$ of 
Dirac fermions due
to logarithmic decrease of  Dirac velocity caused by 
fermion-gauge field interaction.

  We begin with the following continuum effective Lagrangian for 
our problem
\begin{eqnarray}
{\cal L}&=&\bar{\Psi}_{\alpha s}(\partial_{\mu}\gamma^{\mu}+
ia_{\mu}\gamma^{\mu})\Psi_{\alpha s} + \nonumber\\
&&  b^{\ast}(\partial_0-\mu_B+ia_0)b - 
\frac{1}{2m_B}b^{\ast}(\nabla+i {\bf a})^2b.
\label{fb} 
\end{eqnarray}
The Fermi field $\Psi_{\alpha s}$ is a $2\times 1$ spinor:
$\Psi_{1s}^{\dagger}=(f_{1se}^*,f_{1so}^*)$,
$\Psi_{2s}^{\dagger}=(f_{2so}^*,f_{2se}^*)$,
where $\alpha=1,2$ labels the two Fermi points, 
$s=1,..,N$ labels fermion species ($N=2$ for physical
case $s=\uparrow,\downarrow$), and $e,o$ stands for
even and odd sites, respectively.
The $\gamma^{\mu}$  matrices are Pauli
matrices $(\gamma^0,\gamma^1,\gamma^2)=(\sigma^3,\sigma^1,\sigma^2)$ 
and satisfy 
$\{\gamma^{\mu},\gamma^{\nu}\}=2\delta^{\mu\nu}$ ($\mu,\nu=0,1,2$).     
$\bar{\Psi}_{\alpha s}\equiv \Psi_{\alpha s}^{\dagger}\gamma^0$.
In the sF phase of Ref.\cite{wenlee}, the fermion dispersion 
near the fermi points is anisotropic,
 but we  rescale it to an isotropic spectrum $E({\bf k})=v_D|{\bf k}|$ 
where $v_D=\sqrt{v_Fv_2}$, the geometric mean of the 
two velocities ($v_2$
is proportional to the energy gap).
 We set $v_D=1$, unless otherwise specified.
The gauge field $a_{\mu}$ $=(a_0,{\bf a})$ corresponds 
to the $a^3_{\mu}$ part of
the SU(2) gauge fields of Ref.\cite{wenlee}.
The terms in Eq. (\ref{fb}) 
involving the Bose field $b$ (representing charge degree of
freedom) are 
 believed to play several important roles, including
the suppression of chiral symmetry breaking (Neel 
ordering\cite{marston}) and instanton effects\cite{nag}.  
Most importantly, the compressible 
boson current screens the $a_0$ field, making it massive.
Unfortunately  we do not have a detailed  understanding of
our  boson subsystem.  Therefore we shall
draw upon only a few of  qualitative features of the Bose sector
while focusing mainly on the Fermi sector of the theory.
  
Eq. (\ref{fb}) carries certain similarity to the 
uniform resonating valence bond (uRVB) gauge theory\cite{IL,LN} 
 proposed to describe optimally and slighly overdoped cuprates, 
and some of the theoretical framework
 can be carried over
to our problem.  
As in the uRVB case, the internal gauge field $a_{\mu}$
does not have dynamics of its own, but it
 acquires dynamics from the polarization of fermions and bosons.
  Integrating out the
matter fields generates the self energy term for the gauge field
${\cal L}_a =\frac{1}{2}a_{\mu}(\Pi_F^{\mu\nu}+
\Pi_B^{\mu\nu})a_{\nu}$, 
 up to quadratic order. 
The fermion polarization $\Pi_F^{\mu\nu}$  from the 
two Dirac points is  given by
\begin{equation}
\Pi_F^{\mu\nu}(q)=\frac{2N}{\beta}
\sum_{k_0}\!\int\!\!\frac{d^2{\bf k}}{(2\pi)^2}
{\rm tr}\left[G_F(k)\gamma^{\mu}G_F(k+q)\gamma^{\nu}\right], 
\label{pol} 
\end{equation}
where $G_F(k)=-(i k_{\mu}\gamma^{\mu})^{-1}$ is 
the fermion Green's function
and $k,q$ denote 3-momentum; for example, 
$k=(k_0=(2n+1)\pi T,{\bf k})$.  
In the Coulomb gauge, the spatial part and the time part
of the gauge field are decoupled, the propagators being
  $D^{00}(q)=(\Pi_F^{00}(q)+\Pi_B^{00}(q))^{-1}$ and
$D^{ij}(q)=(\delta_{ij}-q_iq_j/{\bf q}^2)D^{\perp}(q)$ 
($i,j=1,2$), with 
$D^{\perp}(q)=(\Pi_F^{\perp}(q)+\Pi_B^{\perp}(q))^{-1}$. 
As mentioned earlier, the
 bosons should  have a finite 
compressibility ($\Pi_B^{00}(q\rightarrow0)\neq 0$) so the  time
component of the gauge field becomes massive (at finite temperature
$\Pi_F^{00}(q\rightarrow0)$ is also nonzero and contributes to the
screening of $a_0$ field), but the spatial part of the gauge field,
which is purely transverse, remains
massless even at finite boson density and temperature, as long
as the bosons are uncondensed (as in the spin gap phase). 
 In the remainder of this paper, we will 
focus on the effect of this massless mode, ignoring the $a_0$ field. 
  
In the absence of detailed understanding of the Bose sector, 
we assume that the transverse gauge propagator 
is dominated by the fermion part. In other words,
$D^{\perp}(q)\approx D_F^{\perp}(q)\equiv 1/\Pi_F^{\perp}(q)$. 
This approximation, which is often used in the uRVB gauge theory,
may not be fully justified in our case, but it allows us to 
organize the infrared behavior of our theory 
within  $1/N$ expansion.
  The full expression for analytically continued 
transverse polarization
function $\Pi_F^{\perp}(\omega,{\bf q})$
at finite temperature
is rather complicated and therefore we shall not write it here, 
although it is used later in the evaluation of
 gauge fluctuation contribution to $\chi_u$ and $c_v^{el}$.  
In the limiting case of  $T> |{\bf q}| >|\omega|$, we have
\begin{equation} 
\Pi_F^{\perp}(\omega,{\bf q})
\approx -iC_1\frac{\omega T}{|{\bf q}|}+C_2\frac{{\bf q}^2}{T},
\label{soft}
\end{equation}
 while in the zero temperature limit,
\begin{eqnarray}
{\rm Im}\Pi_F^{\perp}(\omega,{\bf q})&=& 
-N{\rm sign}(\omega)\theta(|\omega|\!-\!|{\bf q}|)\sqrt{\omega^2-
{\bf q}^2}/8\nonumber\\
{\rm Re}\Pi_F^{\perp}(\omega,{\bf q})&=& N
\theta(|{\bf q}|-|\omega|)\sqrt{{\bf q}^2-
\omega^2}/8.
\end{eqnarray}
  
To the leading order in $1/N$,
 fermion self energy due to transverse
 gauge fluctuations is
\begin{equation}
\Sigma(k)=\frac{1}{\beta}\sum_{q_0}\int 
\frac{d^2{\bf q}}{(2\pi)^2} 
\gamma^i G_F(k+q)\gamma^j D_F^{ij}(q),
\end{equation}
where $D_F^{ij}(q)=(\delta_{ij}-q_iq_j/{\bf q}^2)/\Pi_F^{\perp}$.
At zero temperature, the
 self energy is \cite{diverge}
\begin{eqnarray}
\frac{N}{8}\Sigma(k)= -i\gamma^0\int\!\!\frac{d^3q}{(2\pi)^3}
\frac{k_0+q_0}{(k+q)^2\sqrt{q^2}}\nonumber\\
+i\gamma^x\! \int\!\!\frac{d^3q}{(2\pi)^3}
\frac{(k_x\!+\!q_x)(q_y^2-q_x^2)-
2q_xq_y(k_y\!+\!q_y)}{{\bf q}^2(k+q)^2\sqrt{q^2}}\nonumber \\
+i\gamma^y\! \int\!\!\frac{d^3q}{(2\pi)^3}
\frac{(k_y\!+\!q_y)(q_x^2-q_y^2)-
2q_xq_y(k_x\!+\!q_x)}{{\bf q}^2(k+q)^2\sqrt{q^2}}.
\end{eqnarray}
We find, for $|{\bf k}| > |k_0|$,
\begin{equation}
\Sigma(k)= -c\, ik_0\gamma^0 {\cal A}_0(k)
+ 2c\,i{\bf k}\cdot\vec{\gamma} {\cal A}_1(k)
\end{equation}
with $c=4/(3N\pi^2)$ and
${\cal A}_0(k)\approx {\cal A}_1(k)\approx \ln(\Lambda/|{\bf k}|)$,
where  $\Lambda$ is a UV cutoff.  
Now the pole of the renormalized Green's 
function $G_F^R(k)=(G_F(k)^{-1}-\Sigma(k))^{-1}$ occurs at
\begin{equation}
E({\bf k})= |{\bf k}|(1-4/(N\pi^2) \ln(\Lambda/|{\bf k}|)).
\label{dispersion}
\end{equation}
Note that the presence of compressible bosons and the resulting
 breaking of Lorentz symmetry is crucial to have
logarithmic velocity renormalization.  Indeed, in the 
absense of bosons,
the gauge propagator (gauge independent part) 
is given by $D^{\mu\nu}(q)= 8/N(\delta_{\mu\nu}-
q_{\mu}q_{\nu}/q^2)/\sqrt{q^2}$ and the zero temperature
fermion self energy takes 
the form $\Sigma= ik_{\mu}\gamma^{\mu}f(k^2)$;
therefore the velocity is not renormalized.

Treating the quasiparticles described by Eq. (\ref{dispersion})  
as ``free'',  we calculate
$c_v$ and $\chi_u$  up to $\cal{O}$$(1/N^0)$:
\begin{eqnarray}
c_v^{el}&=&(9/\pi)\zeta(3)N\,
 T^2 (1+(8/N\pi^2)\ln(\Lambda/T) + ..) \nonumber \\
\chi_u&=&(2/\pi)\ln(2)N\,T(1+(8/N\pi^2)\ln(\Lambda/T) + ..),
\label{chiu}
\end{eqnarray}
($\zeta(3)=1.202$).  These results are believed to be valid for
two reasons: 1) ${\rm Im}{\cal A}_{0,1}(\nu+i0^+,{\bf k})=0$ 
for $|\nu|<|{\bf k}|$, so 
the quasiparticles are well-defined.  2)
 Unlike the usual Fermi liquid theory, the free 
particle response function vanishes as $T\rightarrow 0$.  To
the extent the Landau parameters in Fermi liquid theory enter as
in mean field theory, this means that Landau parameter 
correction vanish in $T\rightarrow 0$\cite{leggett}.  
Indeed, it will be shown shortly that
the calculation of $\chi_u$ and $c_v^{el}$ from the 
free energy shift due to gauge fluctuation yields the same results.

The enhancement of $c_v^{el}$ seen here
 finds its counterpart in the more familiar problems such 
as electron-phonon interaction in metals\cite{herring},
uRVB gauge theory\cite{LN}, and half-filled Landau level\cite{hlr},
where interactions induce mass enhancement which manifests iteself
in the  specific heat.  
In the nonrelativistic analogues, however,
mass renormalization does not necessarily result in the 
enhancement of compressibility and uniform 
susceptibility\cite{herring,fukuyama,ybkim},
because  the corrections are tied to the Fermi 
surface\cite{herring}.
The crucial difference in our case is that there are 
only Fermi ``points''
 instead of Fermi ``surface''.   Thus in contrast to the
nonrelativistic case, we find that  the susceptibility is also 
renormalized such that the Wilson ratio $\gamma(T)/\chi_u(T)$ 
is constant. In fact, the Wilson ratio is the
same as that of free Dirac fermions because 
quasiparticles are well-defined
and Fermi-liquid type corrections are absent, as discussed earlier.

To check this conclusion, we calculate
 $\chi_u$ and $c_v^{el}$ in a gauge invariant way,
using the correction to the free energy due to gauge fluctuations. 
 We consider
only the leading correction in $1/N$, 
which is $\cal{O}$$(1/N^0)$:
\begin{equation}
\Delta F\!\!=\!\!\frac{1}{(2\pi)^3}\!\!\int\!\!d^2{\bf q}\!\!
\int_{-\infty}^{\infty}\!\!\!\!\!
d\omega n(\omega)
\tan^{-1}\!\left(\frac{{\rm Im}\Pi_F^{\perp}(\omega,\!{\bf q})}
{{\rm Re}\Pi_F^{\perp}(\omega,\!{\bf q})}\right).
\label{free}
\end{equation}
The entropy shift $\Delta S$ $(=-\partial\Delta F/\partial T)$
 due to gauge fluctuation has two contributions: $\Delta S_1$
from the temperature dependence of the Bose function 
$n(\omega)=1/(\exp(\omega/T)-1)$ and $\Delta S_2$
from the temperature dependence of fermion polarization.  
Numerically we find that the
former gives a $\sim T^2$ contribution to entropy, 
while the latter which can 
be written as 
\begin{equation}
\Delta S_2=\!\!\frac{-1}{(2\pi)^3}\!\!\int^{|{\bf q}|<T_{UV}}
\!\!\!\!\!\!\!\!\!\!\!d^2{\bf q}\!\!\!
\int_{-\infty}^{\infty}\!\!\!
d\omega n(\omega){\rm Im}(D_F^{\perp}
\frac{\partial}{\partial T}\Pi_F^{\perp})
\label{spec}
\end{equation}
($T_{UV}$=high energy cutoff) 
gives a singular contribution  $\propto -T^2 \ln T$.  
The gauge fluctuation contribution to $\chi_u$ $(\Delta\chi_u)$
 is obtained by taking $-\partial^2/\partial H^2$ 
at $H=0$ of $\Delta F(H)$.
This approach
corresponds to summing the bubble diagrams for the vertex 
correction and the self energy correction.   
It takes the form
\begin{equation}
\Delta\chi_u\!\!=\!\!\frac{-1}{(2\pi)^3}\!\!\int^{|{\bf q}|<T_{UV}}
\!\!\!\!\!\!\!\!\!\!\!d^2{\bf q}\!\!\!
\int_{-\infty}^{\infty}\!\!\!
d\omega n(\omega){\rm Im}(D_F^{\perp}
\frac{\partial^2}{\partial \mu_F^2}\tilde{\Pi}_F^{\perp}),
\label{susc}
\end{equation}
where $\partial^2\tilde{\Pi}_{\perp}/\partial \mu_F^2$ 
is a short-hand
notation for $\partial^2\Pi_{\perp}(\omega,{\bf q};\mu_F)/
\partial \mu_F^2|_{\mu_F=0}$ 
in which $\Pi_{\perp}(\omega,{\bf q};\mu_F)$
is the transverse polarization function of Dirac fermions with finite
chemical potential $\mu_F$.  This expression, which closely resembles
that of $\Delta S_2$, gives a singular contribution  $\propto
-T\ln T$.  Note that the expressions for 
$\Delta S_2$ and $\Delta\chi_u$  
are also applicable to (nonrelativistic) uRVB 
gauge theory\cite{LN,hlr},
but they are usually ignored in that case
because they give only  subleading corrections
while $\Delta S_1$ generates a singular correction 
$\propto T^{2/3}$\cite{hlr},
unlike our case in which $\Delta S_2$ dominate at low temperatures.
  Summarizing our numerical evaluation, we have
\begin{equation}
\Delta \chi_u=\frac{0.358}{v_D^2}T\ln \frac{T_{UV}}{2.4T},\,\,\, 
\Delta c_v^{el}= \frac{2.79}{v_D^2}T^2\ln \frac{T_{UV}}{2.6T}
\label{final} 
\end{equation}
at low temperatures ($T<\sim T_{UV}/5$) in agreement
 with Eqs. (\ref{chiu}). 

\narrowtext
\begin{figure}[thb]
\epsfxsize=0.9\columnwidth\epsfbox{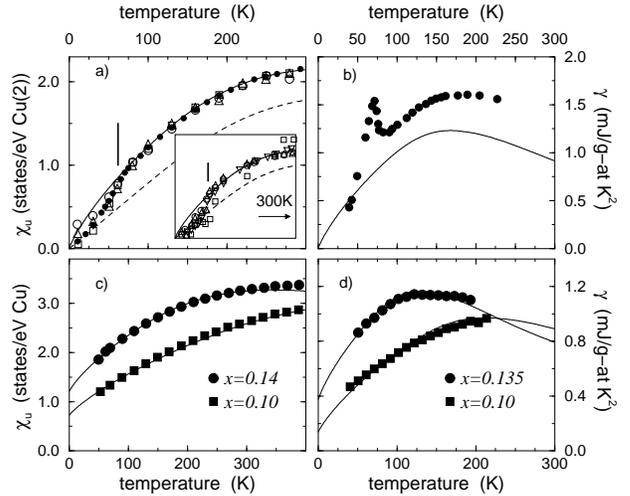}
\caption{ a)$\chi_u$ of   ${\rm YBa_2Cu_3O}_{6.63}$. The inset:
spin Knight shifts of  ${\rm YBa_2Cu_4O}_{8}$.  
The vertical lines indicate $T_c$.
Symbols are as in  Ref.\protect\cite{taki} and \protect\cite{bankay}. 
Dashed line is the susceptibility
$\chi_u^0$ of free Dirac fermions and solid line is the
fit to our thoery, which includes gauge fluctuations.
b)  $\gamma(T)$ of ${\rm YBa_2Cu_3O}_{6.67}$.
c) $\chi_u$ of ${\rm La}_{2-x}{\rm Sr}_x{\rm CuO}_4$ (see text).
 d) $\gamma(T)$ of ${\rm La}_{2-x}{\rm Sr}_x{\rm CuO}_4$.}
\label{unifspec}
\end{figure}
  We now discuss our results in light of
 the experiments.   In Fig. 1a we plot $\chi_u$
 of ${\rm YBa_2Cu_3O_{6.63}}$, a prototypical
underdoped (bilayer) cuprate,  from the Knight shift data of 
Takigawa {\it et al.}\cite{taki}
We took the liberty of moving
 the zero of $\chi_u$  by  0.27 states/eV Cu(2), 
which is within the error bars 
corresponding to uncertainty in the orbital contributions $K^{orb}$
($\chi_u\propto K^{spin}=K-K^{orb}$).  This
change avoids the unphysical situation of 
Ref.\cite{taki} in which
$^{63}\!K^{spin}_{ab},^{17}\!K^{spin}_{iso}, 
^{17}\!K^{spin}_c<0$ at $T=0$. 
 Further support for the adjustment
of 0 comes from precision measurements of the  Knight shifts in   
${\rm YBa_2Cu_4O_8}$ by Brinkmann and collaborators\cite{bankay} 
who made substantial upward
shift of $K^{spin}$ from their previous values\cite{zimmermann}.
We find that the normal state data of
Ref.\cite{taki} are well-fitted (solid line)  by 
$\chi_u(T;v_D,T_{UV})=
\Delta \chi_u+\chi_u^0$.  Here  $\Delta\chi_u$ is the 
numerical evaluation
of Eq. (\ref{susc}) whose low $T$ behavior is 
given by Eq. (\ref{final}),
and  $\chi_u^0$ is the uniform susceptibility 
of bare Dirac fermions 
with the same  upper
cutoff $T_{UV}$: 
$\chi_u^0=\frac{4}{v_D^2\pi}T{\cal F}(T_{UV}/2T)$, 
${\cal F}(x)=\int_0^x y/\cosh^2ydy$.  The two parameters in
the fit are chosen to be 
$v_D=0.76J$ and $T_{UV}=0.63J$, where we set the  antiferromagnetic
exchange energy $J$=1500 K.
We expect the gauge fluctuations to be suppressed 
in the superconducting
state (due to Higgs mechanism) 
 so that $\chi_u$ should cross-over to $\chi_u^0$ (dashed line)
at low temperatures.  This is in qualitative agreement with the data 
below $T_c$.  The inset of
Fig 1a shows a similar fit for the spin Knight shifts of 
${\rm YBa_2Cu_4O_8}$\cite{bankay}, which is again very good.
Thus our theory can account for the susceptibility in both the
normal and superconducting states, without the need to adjust the 
energy scale of the gap parameter.
  {\it Using the same parameters}  $v_D$ and $T_{UV}$
as in the fitting of ${\rm YBa_2Cu_3O_{6.63}}$ Knight shift data, 
we plot $\gamma=\Delta c_v^{el}/T+\gamma^0$ 
(where $\gamma^0=\frac{16}{v_D^2\pi}{\cal G}(T_{UV}/2T),  
{\cal G}(x)=\int_0^x y^3/\cosh^2ydy$) in Fig 1b.
  Also shown is the experimental data for
$\gamma(T)$ of  ${\rm YBa_2Cu_3O}_{6.67}$\cite{loramYBCO}.  
Rough agreement of scales
between the curves is  quite encouraging.

In monolayer   ${\rm La}_{2-x}{\rm Sr}_x{\rm CuO}_4$, 
the uniform susceptibility is
usually deduced from bulk susceptibility by subtracting
the core diagmanetism $\chi_c$ and  Van Vleck paramagnetism 
$\chi_{{\rm vv}}$.  
 Fig. 1c shows  $\chi_u$
of ${\rm La}_{2-x}{\rm Sr}_x{\rm CuO}_4$ 
obtained by subtracting the powder average value 
$\chi_{{\rm vv}}+\chi_c=-0.5$ states/eV\cite{millis} (there's some
uncertainty in the value of $\chi_{{\rm vv}}$) 
from the bulk susceptibility $\chi$\cite{nakano}.  The data
can be characterized  by $\chi_u=\Delta\chi_u+\chi_u^0+\chi_{const}$
 with 
$(v_D=0.99J,T_{UV}=1.17J)$
for $x=0.10$ and $(v_D=0.79J,T_{UV}=0.65J$) for $x=0.14$.  Unlike
the YBCO compounds, temperature independent part 
$\chi_{const}>0$ is
needed for a reasonable fit.  
Regarding  the specific heat data of LSCO, 
 cutoffs significantly smaller than the
ones used for $\chi_u$ are needed to fit 
$\gamma$ of the same compound
in terms of $\gamma=\Delta c_v^{el}/T+\gamma^0+\gamma_{const}$.  
In Fig. 1d we have kept the same $v_D$ as in Fig. 1c,
but used  smaller cutoffs ($T_{UV}=0.8J$ for $x=0.1$ 
and $T_{UV}=0.49J$
for $x=0.135$) to fit the $\gamma$-data\cite{loram}. 
 This  discrepancy 
 and the origin  of nonzero $\gamma_{const}$ and $\chi_{const}$ 
are not well understood.
  The $\chi_{const} > 0$ feature  in 
 LSCO has been emphasized
by some\cite{MIM} to be an important evidence that the 
bilayer structure
is important for spin gap behavior.
Recent experiments on trilayer 
${\rm HgBa_2Ca_2Cu_3O}_{8+\delta}$\cite{julien} and monolayer 
${\rm HgBa_2CuO}_{4+\delta}$\cite{alloul},  
however, find similar spin gap  behaviors  as in YBCO, suggesting
that LSCO is a rather special case.

Despite reasonable agreement, we feel that above comparisons
do not provide
a conclusive test,
because the $T_c$ is too high to probe  the normal state
 infrared behavior 
for a wide range of temperature.  In fact, most of the bending feature
seen in the $\gamma(T)$ data is presumably related to the high energy
cutoff (the deviation from linear Dirac spectrum) which have been treated
in a cavalier manner here by using a hard cutoff $T_{UV}$.  The low-$T$
curvature in $\chi_u$ data ($d^2\chi_u/dT^2<0$; faster decrease at 
lower temperature) seems to support the gauge fluctuation
picture, but it may not be simple to separate this effect from the curvature
due to high energy cutoff.    
Nevertheless, we view that the theory advocated here 
presents a simple and appealing picture of the spin gap phase.  In this
theory, no new energy scale is introduced  to distinguish 
the spin gap phase and the superconducting phase; the Dirac 
velocity in both phases is taken to be the same.
 Rather, it is the gauge fluctuation  that distinguishes
the phases by causing the enhancement of $\chi_u$ and $\gamma$
in the normal state.  

    Instead of conclusion, we recapitulate
 some issues that have been glossed over.
We have ignored the $a_0$ field whose effect may  not be totally 
innocuous\cite{hlubina}.
In fact, we have checked that {\it in the
absence of bosons}
the contributions to $\chi_u$ and $c_v^{el}$ derived from the free
energy shift
due to the $a_0$ fluctuation cancel the singular contributions
from transverse gauge fluctuation,
in agreement with non-renormalization of Dirac velocity in a
Lorentz invariant situation.   Secondly, we have not 
 treated  contributions from the
Bose sector, especially in regard to the entropy.
Lastly, we mention the issue of
 whether the renormalization
of fermion propagator feeds back to the gauge propagator. 
In the non-relativistic gauge theory\cite{LN,hlr}, 
the density-density correlation function and the transverse gauge
propagator receive only sub-leading corrections\cite{ybkim}.  This
might not hold any longer in our case.   
At present it is not clear
 to what extent  the transverse
propagator is modified by ``feedback effect'' and to what extent 
this affects the physical picture.  

DHK acknowledges  helpful conversations with A. Shytov, N. Nagaosa, 
T. Imai, C. Mudry, and Y.B. Kim.
 We have been  supported by 
the NSF MRSEC program [DMR 94-0034] (PAL and DHK)
and by NSF grant No. 94-11574 (XGW).

\end{multicols}
\end{document}